\documentclass[10 pt]{article}

\usepackage{amsmath,amstext,amsgen,amsbsy,amsopn,amsfonts,graphicx,amssymb,
rotating,theorem}

{\theorembodyfont{\rmfamily} \theoremheaderfont{\itshape} 
\newtheorem{thm}{Theorem}}

\begin{document}

\title{Vibrating Quantum Billiards on Riemannian Manifolds}

\author{Mason A. Porter and Richard L. Liboff \\ \\ Center for Applied 
Mathematics \\ and \\ Schools of Electrical Engineering and Applied Physics \\ 
\\ Cornell University}

\date{July, 2000}

\maketitle

\begin{center}
\section*{Abstract}
\end{center}

\vspace{.1 in}

	Quantum billiards provide an excellent forum for the analysis of 
quantum chaos.  Toward this end, we consider quantum billiards with 
time-varying surfaces, which provide an important example of quantum chaos 
that does not require the semiclassical ($\hbar \longrightarrow 0$) or high 
quantum-number limits.  We analyze vibrating quantum billiards using the 
framework of Riemannian geometry.  First, we derive a theorem detailing 
necessary conditions for the existence of chaos in vibrating quantum billiards
 on Riemannian manifolds.  Numerical observations suggest that these 
conditions are also sufficient.  We prove the aforementioned theorem in full 
generality for one degree-of-freedom boundary vibrations and briefly discuss a
 generalization to billiards with two or more degrees-of-vibrations.  The 
requisite conditions are direct consequences of the separability of the 
Helmholtz equation in a given orthogonal coordinate frame, and they arise from
 orthogonality relations satisfied by solutions of the Helmholtz equation.  We
 then state and prove a second theorem that provides a general form for the 
coupled ordinary differential equations that describe quantum billiards with 
one degree-of-vibration boundaries.  This set of equations may be used to 
illustrate KAM theory and also provides a simple example of semiquantum chaos.
  Moreover, vibrating quantum billiards may be used as models for quantum-well
 nanostructures, so this study has both theoretical and practical applications.

\vspace{.3 in}

\subsection*{MSC NOS 37J40, 37K25, 53Z05}

\begin{centering}
\section{Introduction}
\end{centering}

	The study of quantum billiards encompasses an essential subdiscipline 
of applied dynamics.  Within this field, the search for chaotic behavior is 
one component of a large segment of literature concerning quantum 
chaos.\cite{gutz,casati,qc}  The radially vibrating spherical billiard, for 
example, may be used as a model for particle behavior in the 
nucleus\cite{wong} as well as for the quantum dot microdevice 
component.\cite{qdot}.  Additionally, the vibrating cylindrical quantum 
billiard may be used as a model for the quantum wire, another microdevice.  
Other geometries of vibrating quantum billiards have similar applications.  
Moreover, vibrating quantum billiards may be used to model Fermi 
accelerators\cite{fermi,lich}, which provide a description of cosmic ray 
acceleration.  The study of quantum chaos in vibrating quantum billiards 
is thus important both because it expands the mathematical theory of dynamical
 systems and because it can be applied to problems in chemical and mesoscopic 
physics.

	Quantum billiards have been studied extensively in recent 
years.\cite{vibline,gutz,casati,qc}  These systems describe the motion of a 
point particle undergoing perfectly elastic collisions in a bounded domain 
with Dirichlet boundary conditions.  Bl\"umel and Esser\cite{vibline} found 
quantum chaos in the linear vibrating quantum billiard.  Liboff and 
Porter\cite{sazim,sph0} extended these results to spherical quantum billiards 
with vibrating surfaces and derived necessary conditions for these systems to 
exhibit chaotic behavior.  One of the primary goals of this paper is to 
generalize these results to other geometries.

	In the present work, we derive necessary conditions for the existence
 of chaos in vibrating quantum billiards on Riemannian manifolds.  We prove 
such a result in full generality for one degree-of-freedom boundary vibrations
 (henceforth termed \begin{itshape}degree-of-vibration (dov)\end{itshape})
 and also briefly discuss a generalization to quantum billiards with two or
 more \begin{itshape}dov\end{itshape}.  In the ``vibrating quantum billiard 
problem,'' the boundaries of the billiard are permitted to vary with time.  
The \begin{itshape}degree-of-vibration\end{itshape} of the billiard describes 
the number of independent boundary components that vary with time.  If the 
boundary of the billiard is stationary, it is said to have zero 
\begin{itshape}dov\end{itshape}.  The radially vibrating sphere and the linear
 vibrating billiard each have one \begin{itshape}dov\end{itshape}.  The 
rectangular quantum billiard in which both the length and width are 
time-dependent has two \begin{itshape}dov\end{itshape}.

	The requisite conditions for chaotic behavior in one 
\begin{itshape}dov\end{itshape} billiards are direct consequences of the 
separability of the Helmholtz equation\cite{helm} in a given orthogonal 
coordinate frame, and they arise from orthogonality relations satisfied by 
solutions of the Helmholtz equation.  We also state and prove a second theorem
 that gives a general form for the coupled ordinary differential equations 
that describe quantum billiards with one \begin{itshape}dov\end{itshape}.  
These equations provide an illustration of KAM theory, so they are important 
for both research and expository pursuits.

\vspace{.1 in}

\begin{centering}
\section{Quantum Billiards with One Degree-of-Vibration}
\end{centering}

	Quantum billiards describe the motion of a point particle of mass 
$m_0$ undergoing perfectly elastic collisions in a domain in a potential $V$ 
with a boundary of mass $M \gg m_0$.  (Though $m_0/M$ is small, we do not pass
 to the limit in which this ratio vanishes.)  With this condition on the mass 
ratio, we assume that the boundary does not recoil from collisions with the 
point particle confined therein.  Point particles in quantum billiards possess 
wavefunctions that satisfy the Shr\"odinger equation and whose 
time-independent parts satisfy the Helmholtz equation.  Globally separable 
quantum billiards with ``stationary'' (i.e., zero 
\begin{itshape}dov\end{itshape}) boundaries exhibit only integrable behavior. 
 That is, the motion of their associated wavefunctions may only be periodic 
and quasiperiodic.  Two types of quantum billiard systems in which this global
 separability assumption is violated are ones with concave boundary components
 and ones with composite geometry.  Both of these situations exhibit so-called
 ``quantized chaos'' (or ``quantum chaology'').\cite{gutz,atomic}  Perhaps the
 best-known example of a geometrically composite quantum billiard is the 
stadium billiard\cite{katok,arnold,mac}, which consists of two semicircles 
connected by lines to form a `stadium.'  In the present paper, we retain the 
assumption of global separability but permit the boundaries of the quantum 
billiards to vary with time.

\begin{centering}
\subsection{Necessary Conditions for Chaos}
\end{centering}

	In Liboff and Porter\cite{sazim}, it was shown that any 
$k$-superposition state of the radially vibrating spherical quantum billiard 
must include a pair of eigenstates with rotational symmetry (in other words, 
with equal orbital ($l$) and azimuthal ($m$) quantum numbers) in order for the
 superposition to exhibit chaotic behavior.  One of the goals of the present 
paper is to prove the following generalization:

\vspace{.2 in}

\begin{thm}
Let $X$ be an $s$-dimensional Riemannian manifold with (Riemannian) metric 
$g$.  Assume the Helmholtz operator $T \equiv \nabla^2 + \lambda^2$ is 
globally separable on $(X,g)$, so that one may write the wave-function $\psi$ 
as the superposition 
\begin{equation}
	\psi(x) = \sum_{n} \alpha_n(t) A_n(t) \psi_n(x), 
\end{equation}
where $x \equiv (x_1, \cdots, x_s)$ is the position vector, 
$n \equiv (n_1, \cdots, n_s)$ is a vector of quantum numbers, and
\begin{equation} 
	\psi_n(x) = \prod_{j = 1}^s f_j^{(n_j)}(x_j) 
\end{equation}
is a product of $s$ ``component functions'' $f_j^{(n_j)}(x_j)$.  The parameter
 $\alpha_n$ is a normalization constant, and $A_n(t)$ is a complex amplitude.
If the quantum billiard of boundary mass $M$ defined on $(X,g)$ experiences 
one \begin{itshape}dov\end{itshape} in a potential $V = V(a)$, where $a$ 
describes the time-dependent dimension of the boundary, then for any $k$-term 
superposition state to manifest chaotic behavior, it is necessary that there 
exist a pair among the $k$ states whose $s - 1$ quantum numbers not 
corresponding to the vibrating dimension are pairwise equal.  (That is, for 
some pair of eigenstates with respective quantum numbers  $(n_{k_1}, \cdots, 
n_{k_{s - 1}})$ and $(n_{k_1}', \cdots, n_{k_{s - 1}}')$ corresponding to 
non-vibrating dimensions, one must have $n_{k_j} = n_{k_j}'$ for all $j \in 
\{1, \cdots, s - 1\}$.)	
\end{thm}

\vspace{.3 in}

	In words, the above theorem states that given a globally separable 
vibrating billiard, a superposition state of a one 
\begin{itshape}dov\end{itshape} quantum billiard whose geometry is described 
by an $s$-dimensional orthogonal coordinate system must have a pair of 
eigenstates with $(s - 1)$ equal quantum numbers corresponding to the 
stationary dimensions of the billiard's boundary in order to exhibit chaotic 
behavior.  For a discussion of the separability of the Helmholz operator, see 
Appendix I.  Examples of manifolds on which this operator is globally 
separable include well-known ones such as rectangular, cylindrical (polar), 
and spherical coordinates and lesser-known ones such as elliptical cylindrical
 coordinates, parabolic cylindrical coordinates, prolate spheroidal 
coordinates, oblate spheroidal coordinates, and parabolic 
coordinates.\cite{field}  Note that the preceeding list does not exhaust all 
possible coordinate systems.  (There are others in $\mathbb{R}^2$ and 
$\mathbb{R}^3$ and the preeceding examples may be generalized to manifolds in 
higher dimensions for which separability is retained.)  Appendix I includes a 
general procedure for determining if the Helmholz equation is separable for a 
given coordinate system.

	Applying the above theorem to the radially vibrating spherical quantum
 billiard\cite{sazim}, one finds that rotational symmetry between some pair of
 eigenstatesstates in the superposition is required in order for the system to
 exhibit chaotic behavior.  That is, the azimuthal and orbital quantum numbers
 of two of the states in the superposition must be equal.  The value of the 
principal quantum number $n$ does not affect the existence of chaos.

	The solution of the Schr\"odinger equation is of the form\cite{sakurai}
\begin{equation}
	\psi(r,t) = \sum_{n=1}^\infty A_n \alpha_n  e^{-\frac{i E_n t}{\hbar}}
 \psi_n(r).
\end{equation}
Absorbing the resulting time-dependence (in the phase) into the coefficient 
$A_n(t)$ yields
\begin{equation}
	\psi(r,t) = \sum_{n=1}^\infty A_n(t) \alpha_n(t) \psi_n(r,t). 
\end{equation}

	In order to examine the present problem, consider a 2-term 
superposition state of the vibrating billiard in $(X,g)$.  The results for a 
$k$-term superposition state follow from considering the terms pairwise.  
The superposition between the $n$th and $q$th states is given by
\begin{equation}
	\psi_{nq}(x,t) \equiv \alpha_n A_n(t)\psi_n(x,t) + 
\alpha_q A_q(t) \psi_q(x,t). \label{super}
\end{equation}
We substitute this wavefunction into the time-dependent Schr\"odinger 
equation 
\begin{equation}
	i \hbar \frac{\partial \psi(x,t)}{\partial t} = -\frac{\hbar^2}{2m}
\nabla^2 \psi(x,t) , \text{        } x \in X \label{schro},
\end{equation}
where the kinetic energy corresponding to the Hamiltonian of the particle
 confined within the billiard is given by  
\begin{equation}
	K = - \frac{\hbar^2}{2m} \nabla^2.
\end{equation}
The total Hamiltonian of the system is given by
\begin{equation}
	H(a_1, \cdots, a_s, P_1, \cdots, P_s) = K + \sum_{j = 1}^s 
\frac{P_j^2}{2M_j} + V,
\end{equation}
where $a_1, \cdots, a_s$ represent the time-varying boundary components, and 
the walls of the quantum billiard are in a potential $V$ and have momenta 
$P_j$ with corresponding masses $M_j$.  The particle kinetic energy $K$ is the 
quantum-mechanical (fast) component of the Hamiltonian, whereas the remainder 
of the Hamiltonian---representing the potential and kinetic energies of the 
billiard boundary---is the classical (slow) component in this semi-quantum 
system.  We use an adiabatic (Born-Oppenheimer) approximation\cite{vibline} by
 only considering the quantum-mechanical component $K$ of this coupled 
classical-quantum system as the Hamiltonian in the Schr\"odinger equation.  
The Born-Oppenheimer approximation is commonly used in mesoscopic physics.  In
 this analysis, we also neglect Berry phase.\cite{berry}

	For the present configuration, we assume that V does not depend 
explicitly on time.  That is,
\begin{equation}
	V = V(a_1, \cdots, a_s).
\end{equation}
Note that we are applying nonlinear boundary conditions:
\begin{equation}
	\psi(a_1(t), \cdots, a_s(t)) = 0. \label{bcs}
\end{equation}

	Taking the expectation of both sides of (\ref{schro}) for the state 
(\ref{super}) gives the following relations:
\begin{gather}
	\left\langle \psi_{nq} \left| - \frac{\hbar^2}{2m}\nabla^2 \psi_{nq} 
\right. \right\rangle = K(|A_n|^2,|A_q|^2,a_1, \cdots, a_s) \notag \\
	i \hbar \left\langle \psi_{nq} \left|  \frac{\partial \psi_{nq}}
{\partial t}\right. \right\rangle = i\hbar [\dot{A}_n A_q^* + \dot{A}_q 
A_n^* + \nu_{nn} |A_n|^2 + \nu_{qq} |A_q|^2 + \nu_{nq} A_n A_q^* + \nu_{qn} 
A_q A_n^*]. \label{expect}
\end{gather}

	In a one \begin{itshape}dov\end{itshape} billiard, $a(t) \equiv 
a_1(t)$ is the only time-dependent boundary term (with corresponding momentum 
$P(t) \equiv P_1(t)$), so in this case, the kinetic
 energy is written 
\begin{equation}
	K = K(|A_1|^2,|A_2|^2,a),  
\end{equation}
where we use the notation $A_1 \equiv A_n$, $A_2 \equiv A_q$.  The 
potential energy is given by
\begin{equation}
	V = V(a).
\end{equation}
In this case, there is a single momentum term in $H$ given by $P^2/2M.$  
Liboff and Porter\cite{sazim} showed for the radially vibrating spherical 
billiard that if $\psi_n$ and $\psi_q$ do not have common angular-momentum 
quantum numbers, then $\mu_{nq} = \mu_{qn} = 0$, where the coupling 
coefficient $\mu_{nn'}$ is defined by 
\begin{equation}
	\nu_{nn'} \equiv \mu_{nn'}\frac{\dot{a}}{a}.  
\end{equation}
We show that the vanishing of these coefficients in any one 
\begin{itshape}dov\end{itshape} quantum billiard implies non-chaotic behavior 
of a given superposition state of that billiard.  Without such cross terms, 
one observes that $\dot{A}_j$ is a function of only $A_j$ and $a$:
\begin{equation}
	\dot{A}_j = \chi_j(A_j,a).
\end{equation}
Therefore, $|A_j(t)|^2$ is a function only of $a(t)$, and so the present 
system has the Hamiltonian 
\begin{equation}
	H(a,P) = K(a) + \frac{P^2}{2M} + V(a), \label{ham}
\end{equation}
where 
\begin{equation}
	P = -i \hbar \nabla_a  \equiv  \hat{a} \cdot \nabla  \equiv  -i \hbar 
\frac{\partial}{\partial a}
\end{equation}
is the momentum of the billiard's boundary.  The symbol $\nabla_a$ represents 
the component of the gradient in the direction $\hat{a}$.  In spherical 
coordinates, for example, we identify $\hat{a}$ with the unit vector in the 
radial direction, and thus $\nabla_a$ represents the component of the gradient 
in the radial direction.

	The Hamiltonian (\ref{ham}) describes an autonomous single 
degree-of-freedom system, which corresponds to a 2-dimensional autonomous
 dynamical system, whose non-chaotic properties are 
well-established.\cite{gucken, wiggins}  We therefore conclude that at least 
one of the coupling coefficients $\mu_{nq}$ or $\mu_{qn}$ must be non-zero for
 a quantum billiard with one \begin{itshape}dov\end{itshape} to exhibit 
chaotic behavior.  We show below for separable systems (see Appendix I) that 
the condition for the coupling coefficients $\mu_{nq}$ to vanish is a 
consequence of the orthogonality of the superposition's component functions 
$f_j(x_j)$ (see equation (\ref{sep}) in Appendix I).
	
	In the case of a $k$-term superposition, one considers the coupling 
coefficients $\{\mu_{nq}\}$ of each pair of eigenstates in the superposition. 
 If any one of these coupling terms is non-zero, then one expects the system 
to exhibit chaotic behavior.  Indeed, the fact that the coupling coefficients 
do not vanish implies that one obtains a 5-dimensional dynamical system (which
 is really a two degree-of-freedom Hamiltonian system in disguise).  One 
observes numerically that no matter which two terms one considers, the 
resulting dynamical system behaves chaotically for some set of parameters and 
initial conditions.  Note that the above theorem does not hold for two 
\begin{itshape}dov\end{itshape} quantum billiards, because if one considers an
 inseparable potential such as the anharmonic potential, then one has a two 
degree-of-freedom Hamiltonian system even for cases in which the coupling 
coefficient vanishes.  In the next subsection, we discuss the technical 
details of the proof of this theorem.

\vspace{.1 in}

\begin{centering}
\subsection{Orthogonality of the Component Functions}
\end{centering}
	
	Using the method of separation of variables (again see equation 
(\ref{sep}) in Appendix I) on the Helmholtz equation for a system with $s$ 
degrees-of-freedom, one obtains $s$ boundary-value problems to solve.  (Note 
that the ordinary differential equations for $f_j$ are Sturm-Liouville 
problems.)  The solutions to such problems may be expressed as eigenfunction 
expansions \cite{butkov}.  The orthogonality properties of the resulting 
eigenfunctions are well-known.  For each $j$,
\begin{equation}
	\left\langle f_j^{n_j} \left. \right | f_j^{n_j'} \right\rangle = 
\delta_{n_j n_j'}. \label{delta}
\end{equation} 

	When taking the expectation of the right side of equation 
(\ref{super}), the orthogonality relations satisfied by the $s$ component 
functions $f_j$ play an essential role.  In the following discussion, 
\begin{itshape}fb\end{itshape} and \begin{itshape}mb\end{itshape} denote fixed
 boundaries and movable boundaries, respectively.  When calculating the 
expectation, one must integrate with respect to all $s$ variables to see when 
the inner product in (\ref{delta}) is non-zero.  In particular, this inner 
product is present in each of the cross terms for the $(s - 1)$ 
\begin{itshape}fb\end{itshape} variables, so those terms vanish unless $n_j = 
n_j'$ for each of these $(s - 1)$ variables.  By the separability of $\psi$, 
the $s$-dimensional expectation integral is expressible as the product of $s$ 
one-dimensional integrals, so each term includes a prefactor that consists of 
the product of $(s - 1)$ inner products.  Using the Chain Rule, one finds that
 a variable whose corresponding boundary is time-dependent 
(``\begin{itshape}mb\end{itshape} variables'') will manifest differently in 
the calculation.\cite{sazim}  Since the \begin{itshape}fb\end{itshape} 
variables must have corresponding symmetric \begin{itshape}fb\end{itshape} 
quantum numbers ($n_j = n_j'$ for all $j \in \{k_1, \cdots, k_{s - 1}\}$) for 
a two-state superposition to have non-zero coupling coefficients 
$\{\mu_{nq}\}$, and since we showed above that there must be at least one such
 cross term for a one \begin{itshape}dov\end{itshape} quantum billiard to 
exhibit chaotic behavior, there must exist a pair of eigenstates whose 
$(s - 1)$ \begin{itshape}fb\end{itshape} quantum numbers are equal.  This 
completes the proof of Theorem 1.

\vspace{.1 in}

\begin{centering}
\section{Quantum Billiards with Two or More Degrees-of-Vibration}
\end{centering}

	We now generalize the above results to quantum billiards with 
vibrations of two or more degrees-of-freedom.  Suppose that $\xi$ of 
the $s$ boundary components are time-dependent and also suppose that the 
Hamiltonian is separable:
\begin{equation}
	H(a_1, \cdots, a_\xi, P_1, \cdots, P_\xi) = 
\sum_{j = 1}^\xi H_j(a_j,P_j).
\end{equation}
	For $H$ to be separable, one requires that both the billiard's 
potential $V(a_1, \cdots, a_\xi)$ and the kinetic energy 
$K(|A_1|^2, \cdots , |A_k|^2 , a_1, \cdots, a_\xi)$ be separable in the same 
sense as the Hamiltonian.  For some systems, such as the vibrating rectangular
 parallelepiped, the kinetic energy is separable.  For others, this need not 
be the case.  For example, the spherical billiard has kinetic energy 
$K(r,\theta,\phi) = K_1(r)K_2(\theta,\phi)$, which is not equal to 
$K_1 + K_2$.  

	If, in a given superposition, there are no cross terms in the 
expectation (\ref{expect}$'$), the $\xi$ degree-of-freedom autonomous 
Hamiltonian above gives rise to a $2\xi$-dimensional autonomous system, which,
 because of the separability, decouples into $\xi$ two-dimensional autonomous 
systems, whose non-chaotic properties are well-known.  If either $V$ or $K$ is 
not separable, the system does not decouple.  One therefore cannot conclude 
that the system does not have chaotic behavior even in the absense of cross 
terms.  A given system is very likely to behave chaotically in this event.  In
 the separable case, then, a $k$-term superposition state exhibits 
chaotic behavior when the corresponding $(s - \xi)$ 
\begin{itshape}fb\end{itshape} quantum numbers must be the same for some pair 
of eigenstates (that is, the $i$th \begin{itshape}fb\end{itshape} quantum 
number in one state must be the same as the $i$th 
\begin{itshape}fb\end{itshape} quantum number in the other state of the pair. 
 Here, $i$ runs over all $(s - \xi)$ \begin{itshape}fb\end{itshape} quantum 
numbers).  The proof is entirely analogous to the one above.

\vspace{.1 in}

\begin{centering}
\section{Differential Equations for One Degree-of-Vibration Quantum Billiards}
\end{centering}

	In the present section, we derive a general form for the coupled 
differential equations describing quantum billiards (in a separable coordinate
 system) with one \begin{itshape}dov\end{itshape} and a nonvanishing coupling 
coefficient $\mu_{nq}$.  The resulting system of ordinary differential 
equations behaves chaotically.  Indeed, numerical simulations indicate chaotic
 behavior for some choices of initial conditions and parameters.

\vspace{.2 in}

\begin{thm}
Consider a one \begin{itshape}dov\end{itshape} quantum billiard with the 
same geometric conditions as in Theorem 1.  Let the point particle inside the 
billiard be of mass $m_0$, the mass of the billiard's boundary be $M \gg m_0$,
 and the surface potential of the billiard be $V = V(a)$, where $a = a(t)$.  
For a two-term superposition, if the $i$th \begin{itshape}fb\end{itshape} 
quantum number is the same in both states [where $i$ runs over all $(s - 1)$
 of these numbers], then the system of differential equations describing the 
evolution of the superposition state has the following form in terms of Bloch 
variables $x$, $y$, $z$ (defined below)\cite{bloch}, displacement $a$, 
and momentum $P$:
\begin{equation}
	\dot{x} = -\frac{\omega_0 y}{a^2} - \frac{2 \mu_{qq'} P z}{Ma}, 
\label{xdot}
\end{equation}
\begin{equation}
	\dot{y} = \frac{\omega_0 x}{a^2},
\end{equation}
\begin{equation}
	\dot{z} = \frac{2 \mu_{qq'} P x}{Ma},
\end{equation}
\begin{equation}
	\dot{a} = \frac{P}{M}, 
\end{equation}
and
\begin{equation}
	\dot{P} = -\frac{\partial V}{\partial a} + \frac{2[\epsilon_+ + 
\epsilon_-(z - \mu_{qq'}x)]}{a^3}. \label{Pdot}
\end{equation}

	In the above equations, 
\begin{equation}
	\omega_0 \equiv \frac{\epsilon_{q'} - \epsilon_q}{\hbar},
\end{equation}
and
\begin{equation}
	\epsilon_{\pm} \equiv \frac{(\epsilon_{q'} \pm \epsilon_q)}{2},
\end{equation}
where $\epsilon_q$ and $\epsilon_{q'}$ $(q \neq q')$ are the coefficients in 
the kinetic energy corresponding to the \begin{itshape}mb\end{itshape} quantum
 numbers.  Additionally, $x$, $y$, and $z$ represent (dimensionless) Bloch 
variables:  
\begin{gather}
	x = \rho_{12} + \rho_{21}, \label{blochvars} \\ 
	y = i(\rho_{21} - \rho_{12}), \tag{\ref{blochvars}$'$} \\ 
	z = \rho_{22} - \rho_{11}, \tag{\ref{blochvars}$''$}
\end{gather}
where the density matrix is defined by $\rho_{qn} = A_q A_n^*$.\cite{liboff}
\end{thm}

\vspace{.3 in}

	Before we begin the proof of Theorem 2, note that the differential 
equations describing the evolution of a two-term superposition state are of 
the above form for the linear vibrating billiard \cite{vibline,atomic} as well
 as for the vibrating spherical billiard with both vanishing and non-vanishing
 angular momentum eigenstates.\cite{sazim,sph0}  Recall that this system of 
equations has two constants of motion.  They are the energy (Hamiltonian)
\begin{equation}
	H = \text{constant}
\end{equation}
and the radius of the Bloch sphere
\begin{equation}
	x^2 + y^2 + z^2 = 1,
\end{equation}
so there are three independent dynamical variables in the set 
$\{x,y,z,a,P\}$.

	We verify Theorem 2 using techniques from Riemannian geometry.  It is 
well-known that for $s$-dimensional Riemannian manifolds, the volume element 
$dV$ has units of (distance)$^s$, which may include some ``prefactors.''  (For
 example, in cylindrical coordinates, $dV = r dr d\theta dz$, where $r$ is the
 prefactor.)  In particular, if there are $\xi$ ``angular variables''
 (dimensionless quantities, like $\theta$ in the above example), there will be
 a prefactor that includes the term $r^\xi$ so that the volume element
 has appropriate dimensions.  Additionally, in a quantum billiard with one 
\begin{itshape}dov\end{itshape} corresponding to the boundary dimension 
$a(t)$, the wave-function has a normalization factor of order 
$a^{-\frac{\sigma}{2}}$, where $\sigma$ corresponds to the number of distance 
dimensions affected by the vibration, which is a different concept from the 
\begin{itshape}dov\end{itshape}.  For example, in the radially vibrating 
sphere, the vibration of the radius affects three dimensions, even though this
 system has one \begin{itshape}dov\end{itshape}.  In contrast, for a rectangle 
in which either the length or width (but not both) is time-dependent, a single
 distance dimension is affected, and the \begin{itshape}dov\end{itshape} is 
also one.  When taking the expectation of the Schr\"odinger equation 
(\ref{schro}), the normalization prefactor of $\psi$ is squared, which gives a
 factor of $a^{-\sigma}$.  We perform $s$ inner products(and hence $s$ 
integrations) in taking this expectation, which yields a factor of 
$\dot{a}/a$ in each of the cross terms, as was the case for known 
examples.\cite{riemann,mta}  The diagonal terms in the expectation 
(\ref{expect}) are due only to the kinetic energy, because of orthogonality 
conditions on the wavefunctions $\psi_q$ and $\psi_{q'}$.

	The evolution equations for $\dot{A}_1$ and $\dot{A}_2$ (see 
(\ref{super}), \begin{itshape}etc.\end{itshape}) are thus
\begin{equation}
	i \dot{A}_n = \sum_{j=1}^2 D_{nj} A_j, \label{amp}
\end{equation}
where
\begin{equation}
	(D_{nj}) =  
	\begin{pmatrix}
		\frac{\epsilon_q}{\hbar a^2} & -i \mu \frac{\dot{a}}{a} \\
		i \mu \frac{\dot{a}}{a} & \frac{\epsilon_{q'}}{\hbar a^2}
	\end{pmatrix},
\end{equation}	
and $\mu \equiv \mu_{qq'} = -\mu_{q'q} \neq 0$ is a coupling coefficient 
(proportional to $\nu_{qq'}$) for the cross term $A_q A_{q'}^*$ corresponding 
to equation (\ref{expect}).  Transforming these amplitudes to Bloch variables 
(\ref{blochvars}) completes the proof of Theorem 2.  The calculation is 
exactly as in the radially vibrating spherical billiard.\cite{sazim,sph0}  

	The above equations may be used to illustrate KAM 
theory.\cite{gucken,wiggins,katok}  Toward that purpose, the number of 
nonresonant tori that have broken up depends on the initial condition of a 
given integral curve.  One may obtain, for example, periodic and quasiperiodic 
orbits (corresponding to closed curves in the Poincar\'e map) as in Figures 1 
and 2, local (``soft'') chaos (in which these closed curves become ``fuzzy'') 
as in Figure 3, structured global chaos (Figure 4), islands of order in a sea 
of chaos (Figure 5), and finally global chaos (Figure 6).  The Poincar\'e 
sections corresponding to the descriptions above for the evolution equations 
of a one \begin{itshape}dov\end{itshape} quantum billiard have initial 
conditions and parameter values  given by $x(0) = \sin(0.95 \pi) \approx 
0.15643446504$, $y(0) = 0$, $z(0) = 
\cos(0.95 \pi) \approx -0.987688340595$, $\frac{V_0}{a_0^2} = 5$, $a_0 = 
1.25$, $\hbar = 1$, $\epsilon_1 = 5$, $\epsilon_2 = 10$, and $\mu = 1.5$.  
Figures 1--6 are plots for the harmonic potential
\begin{equation}
	V = \frac{V_0}{a_0^2}(a - a_0)^2.
\end{equation}

\begin{centering}
\section{Phenomenology}
\end{centering}

	We now discuss the phenomenology of quantum chaos in the present 
context.  In the language of Bl\"umel and Reinhardt\cite{atomic}, vibrating 
quantum billiards are an example semiquantum chaos, which describes different
 behavior than the so-called ``quantized chaos'' that is more commonly 
studied.  Quantized chaos or ``quantum chaology'' is the study of the quantum 
signatures of classically chaotic systems, usually in the semiclassical 
($\hbar \longrightarrow 0$) or high quantum-number limits.  The observed 
behavior in these studies is not strictly chaotic, but the non-intgrability of
 these systems is nevertheless evident.  The fact that their classical analogs
 are genuinely chaotic has a notable effect on the quantum dynamics.\cite{qbc}
  In the semiquantum chaotic situation, the semiclassical and high 
quantum-number limits are unnecessary and the observed behavior is genuinely 
chaotic.

	In vibrating quantum billiards, one has a classical system (the walls 
of the billiard) coupled to a quantum-mechanical one (the particle enclosed by
 the billiard boundary).  Considered individually, each of these subsystems is
 integrable.  When they are coupled, however, one observes chaotic behavior in
 each of them.  (Physically, the coupling occurs when the particle confined 
within the billiard strikes the vibrating boundary.  The motions of the 
particle and wall thereby affect each other.)  The classical variables $(a,P)$
 exhibit Hamiltonian chaos, whereas the quantum subsystem $(x,y,z)$ is truly 
quantum chaotic.  Chaos on the Bloch sphere is an example of quantum chaos, 
because the Bloch variables $(x,y,z)$ correspond to the quantum probabilities 
of the wavefunction.  Additionally, a single normal mode depends on the radius
 $a(t)$, and so each eigenfunction is an example of quantum-mechanical wave 
chaos for the chaotic configurations of the billiard.  Because the evolution 
of the probabilities $|A_i|^2$ is chaotic, the waevfunction $\psi$ in the 
present configuration is a chaotic combination of chaotic normal modes.  This 
is clearly a manifestation of quantum chaos.  Finally, we note that if we 
quantized the motion of the billiard's walls, we would obtain a 
higher-dimensional, fully-quantized system that exhibits quantized 
chaos.\cite{atomic}  In particular, the fully quantized version of the present
 system would require passage to the semiclassical limit in order to observe 
quantum signatures of classical chaos.

\begin{centering}
\section{Conclusions}
\end{centering}

\vspace{-.05 in}

	We derived necessary conditions for the existence of chaos in one 
degree-of-vibration quantum billiards on Riemannian manifolds (Theorem 1).  
In a $k$-state superposition, there must exist a pair of states whose $fb$ 
quantum numbers are pairwise equal.  The results of this theorem arise from 
the separability of the Schr\"odinger equation for a given orthogonal 
coordinate system as well as orthogonality relations satisfied by solutions of
 the Schr\"odinger equation.  We also discussed a generalization of the 
previous result to vibrating quantum billiards with two or more
\begin{itshape}dov\end{itshape}.  Moreover, we derived a general form 
(Theorem 2) for the coupled equations that describe vibrating quantum 
billiards with one \begin{itshape}dov\end{itshape}, and we used these 
equations to illustrate KAM theory.  

	We showed that the equations of motion (\ref{xdot}--\ref{Pdot}) for
 a one \begin{itshape}dov\end{itshape} quantum billiard describe a 
class of problems that exhibit semiquantum chaotic 
behavior.\cite{vibline,atomic,sazim,sph0}  Unlike in quantum chaology, the 
behavior in question is genuinely chaotic.  Additionally, we did not need to 
pass to the semi-classical ($\hbar \longrightarrow 0$) or high quantum-number 
limits in order to observe such behavior.  From a more practical standpoint, 
the radially vibrating spherical billiard may be used as a model for particle 
behavior in the nucleus\cite{wong}, the `quantum dot' 
nanostructure\cite{qdot}, and the Fermi accelerating sphere\cite{fermi}.  The 
vibrating cylindrical billiard may be used as a model of the `quantum wire' 
microdevice component.\cite{liboff,qwire}  At low temperatures, quantum-well 
nanostructures experience vibrations due to zero-point motions, and at high 
temperatures, they vibrate because of natural fluctuations.  Additionally, the
 `liquid drop' and `collective' models of the nucleus include boundary 
vibrations.\cite{wong}  The present paper thus has both theoretical and 
practical import because it expands the mathematical theory of quantum chaos 
and has application in nuclear, atomic, and mesoscopic physics.

\vspace{.1 in}

\begin{centering}
\section*{Appendix I: Separability of the Helmholtz Operator}
\end{centering}

	Consider an $s$-dimensional Riemannian manifold with metric 
coefficients $\{g_{11}, \cdots, g_{ss}\}$ defined by
\begin{equation}
	g_{jj} = \sum_{i = 1}^s \left(\frac{\partial x_i}
{\partial u_j}\right)^2,
\end{equation}
where $x_i$ represents the $i$th rectangular coordinate and $u_j$ represents 
the distance along the $j$th axis \cite{crc}.  The Riemannian metric is then 
$g = \prod_{j = 1}^s g_{jj}$.  For notational convenience, one defines $h_j = 
\sqrt{g_{jj}}$, so that $\sqrt{g} = \prod_{j = 1}^s h_j$.  In cylindrical 
coordinates in $\mathbb{R}^3$, for example, $x_1 = r \text{cos}(\theta)$, 
$x_2 = r \text{sin}(\theta)$, and $x_3 = z$, so that one obtains $h_1 = 1$, 
$h_2 = r$, and $h_3 = 1$.

	We review the following analysis from Riemannian geometry so that the 
proof of Theorem 1 is easier to follow.  To express the Helmholtz equation on 
$(X,g)$, one writes the Laplace-Beltrami operator $\nabla^2$ with respect to 
the metric $g$:
\begin{equation}
	\nabla^2 = \frac{1}{\sqrt{g}}\sum_{j = 1}^s  \frac{\partial}{\partial 
u_j}\left(\frac{\sqrt{g}}{g_{jj}} \frac{\partial}{\partial u_j}\right).	
\end{equation}
If the manifold $X$ is three-dimensional, the Laplacian takes the form
\begin{gather}
	\nabla^2 = \frac{1}{\sqrt{g}} \left[ \frac{\partial}{\partial u_1} 
\left(\frac{\sqrt{g}}{g_{11}} \frac{\partial}{\partial u_1}\right) + 
\frac{\partial}{\partial u_2}\left(\frac{\sqrt{g}}{g_{22}} \frac{\partial}
{\partial u_2}\right) + \frac{\partial}{\partial u_3}\left(\frac{\sqrt{g}}
{g_{33}} \frac{\partial}{\partial u_3}\right) \right] \notag \\
	= \frac{1}{h_1 h_2 h_3} \left[ \frac{\partial}{\partial u_1} 
\left(\frac{h_2 h_3}{h_1} \frac{\partial}{\partial u_1}\right) + 
\frac{\partial}{\partial u_2}\left(\frac{h_3 h_1}{h_2} \frac{\partial}
{\partial u_2}\right) + \frac{\partial}{\partial u_3}\left(\frac{h_1 h_2}
{h_3} \frac{\partial}{\partial u_3}\right) \right] \label{lap}
\end{gather}

	We now discuss the separability of the Helmholtz equation $\nabla^2 
\psi + \lambda^2 \psi = 0$, which is one of our geometrical conditions.  To do 
so, define the St\"ackel matrix\cite{stackel,field} 
\begin{equation}
	S \equiv (\Phi_{ij}),
\end{equation}
	where $\Phi_{ij} = \Phi_{ij}(u_i)$, and the $\{\Phi_{ij}\}$ are 
specified by the following procedure.  Define
\begin{equation}
	C \equiv \text{det}(S) = \sum_{j = 1}^s \Phi_{j1} M_{j1}.
\end{equation}
In the preceeding equation, the $(j,1)$-cofactor $M_{j1}$ is given by 
\begin{equation}
	M_{j1}  = (-1)^{j + 1} \text{det}\left[M(j|1)\right],
\end{equation}
where $M(j|i)$ represents the $(j,i)$-cofactor matrix that one obtains by 
considering the submatrix of $S$ defined by deleting the $j$th row and the 
$i$th column \cite{strang}.  In three dimensions, $M_{j1}$ take the form
\begin{equation}
	M_{11} = \begin{vmatrix}
			\Phi_{22} & \Phi_{23} \\ \Phi_{32} & \Phi_{33}
		 \end{vmatrix},
\end{equation}
\begin{equation}
	-M_{21} = \begin{vmatrix}
			\Phi_{12} & \Phi_{13} \\ \Phi_{32} & \Phi_{33}
		  \end{vmatrix}, 
\end{equation}
and
\begin{equation}
	M_{31} = \begin{vmatrix}
			\Phi_{12} & \Phi_{13} \\ \Phi_{22} & \Phi_{23}
		 \end{vmatrix}.
\end{equation}

If 
\begin{equation}
	g_{jj} = \frac{C}{M_{j1}}
\end{equation}
and
\begin{equation}
	\frac{\sqrt{g}}{C} = \prod_{j = 1}^s \eta_j(u_j),
\end{equation}
then the solution of the Helmholtz equation 
\begin{equation}
	\nabla^2 \psi + \lambda^2 \psi = 0 \label{helmet}
\end{equation}
separates:
\begin{equation}
	\psi = \prod_{j = 1}^s f_j(u_j), \label{sep}
\end{equation}
where $f_j$ solves the Sturm-Liouville equation\cite{simmons}
\begin{equation}
	\frac{1}{\eta_j}\frac{d}{d u_j}\left(\eta_j \frac{d f_j}{d u_j}\right)
 + f_j\sum_{i = 1}^s b_i \Phi_{ji} = 0, \text{        }j \in \{1, \cdots, s\} 
\label{fj}
\end{equation}

	In (\ref{fj}), $b_1 = \lambda^2$, and all other $b_i$ are 
arbitrary.  For a given St\"ackel matrix, this prescribes the $\{\eta_j\}$ for
 which the separability conditions hold.  It is important to note that for a 
given metric $g$, the choice of the St\"ackel matrix is not unique and that 
for some metrics, there is no St\"ackel matrix that can be chosen and hence 
no way to separate the Helmholtz operator. Note also that the time-dependent 
Schr\"odinger equation is separable whenever the Helmholtz equation is 
separable.

	As a special case [corresponding to $\lambda = 0$ in (\ref{helmet})],
Laplace's equation is separable whenever
\begin{equation}
	\frac{g_{jj}}{g_{kk}} = \frac{M_{k1}}{M_{j1}} 
\end{equation}
and
\begin{equation}
	\frac{\sqrt{g}}{g_{jj}} = M_{j1} \prod_{i = 1}^s \eta_i(u_i), 
\hspace{0.5 in} j,k \in \{1, \cdots, s\}.
\end{equation}	
Note that the preceeding condition does not completely decribe the 
separability of the Helmholtz operator, because although the Helmholtz 
equation is separable whenever the Laplacian is separable, the converse is not 
true.

\vspace{.6 in}

\begin{centering}
\section*{Appendix II: Gal\"erkin Approximations}
\end{centering}

\vspace{-.05 in}

	The method used to obtain nonlinear coupled ordinary differential 
equations for the amplitudes $A_j$ amounts to applying the Gal\"erkin 
method\cite{gucken,infinite} to the Schr\"odinger equation, an 
infinite-dimensional dynamical system.  It has been used for many years to 
study nonlinear reaction-diffusion equations that occur in fluid mechanics. 
 It can also be used in the study of nonlinear Schr\"odinger equations (NLS).
  Our treatment of the 
linear Schr\"odinger equation with nonlinear boundary conditions thus 
parallels established methods for nonlinear partial differential equations.  
Additionally, the Finite Element Method is also based on a Gal\"erkin 
approximation\cite{fem} and one can use such methods in inertial manifold 
theory.

	The Gal\"erkin method proceeds as follows.  Consider a partial 
differential equation (possibly nonlinear) whose solution is the function 
$\psi$.  Express $\psi$ as an expansion in some orthonormal set of 
eigenfunctions $\psi_i(x), i \in I$:
\begin{equation}
	\psi(x) = \sum_I c_i(\bar{x}) \psi_i(x), \hspace{.2 in} x \in X, 
\label{gal}
\end{equation}
where $I$ is any indexing set and the coefficients $c_i(\bar{x})$ are unknown 
functions of some but not all of the independent variables in the vector $x$, 
as in the present paper.  This gives a countably infinite coupled system of 
nonlinear ordinary differential equations for $c_i(\bar{x}), i \in I$.  (If 
the partial differential equation is linear with linear boundary conditions, 
then taking an eigenfunction expansion gives constant coefficients 
$c_i(\bar{x}) \equiv c_i$.)  One then projects the expansion (\ref{gal}) onto 
a finite-dimensional space (by assuming that only a certain finite subset of 
the $c_i(\bar{x})$ are non-zero) to obtain a finite system of coupled ordinary 
differential equations.  Thus, a two-term superposition state corresponds to a
 two-term Gal\"erkin projection.  If all the dynamical behavior of a system 
lies on such a finite-dimensional projection, then one has found an inertial 
manifold of the system.\cite{infinite}

\begin{centering}
\section*{Acknowledgements}
\end{centering}

	We would like to thank Vito Dai, Adrian Mariano, and Catherine Sulem 
for fruitful discussions concerning this topic.

\vspace{.1 in}

\bibliographystyle{plain}
\bibliography{ref}

\vspace{.5 in}

\begin{centering}
\subsection*{Figure Captions}
\end{centering}

\vspace{.1 in}

Figure 1: Periodicity and Quasiperiodicity I in a one $dov$ quantum billiard

\vspace{.1 in}

Figure 2: Periodicity and Quasiperiodicity II in a one $dov$ quantum billiard

\vspace{.1 in}

Figure 3: Local Chaos in a one $dov$ quantum billiard

\vspace{.1 in}

Figure 4: Structured Global Chaos in a one $dov$ quantum billiard

\vspace{.1 in}

Figure 5: Islands in a Sea of Chaos in a one $dov$ quantum billiard

\vspace{.1 in}

Figure 6: Global Chaos in a one $dov$ quantum billiard

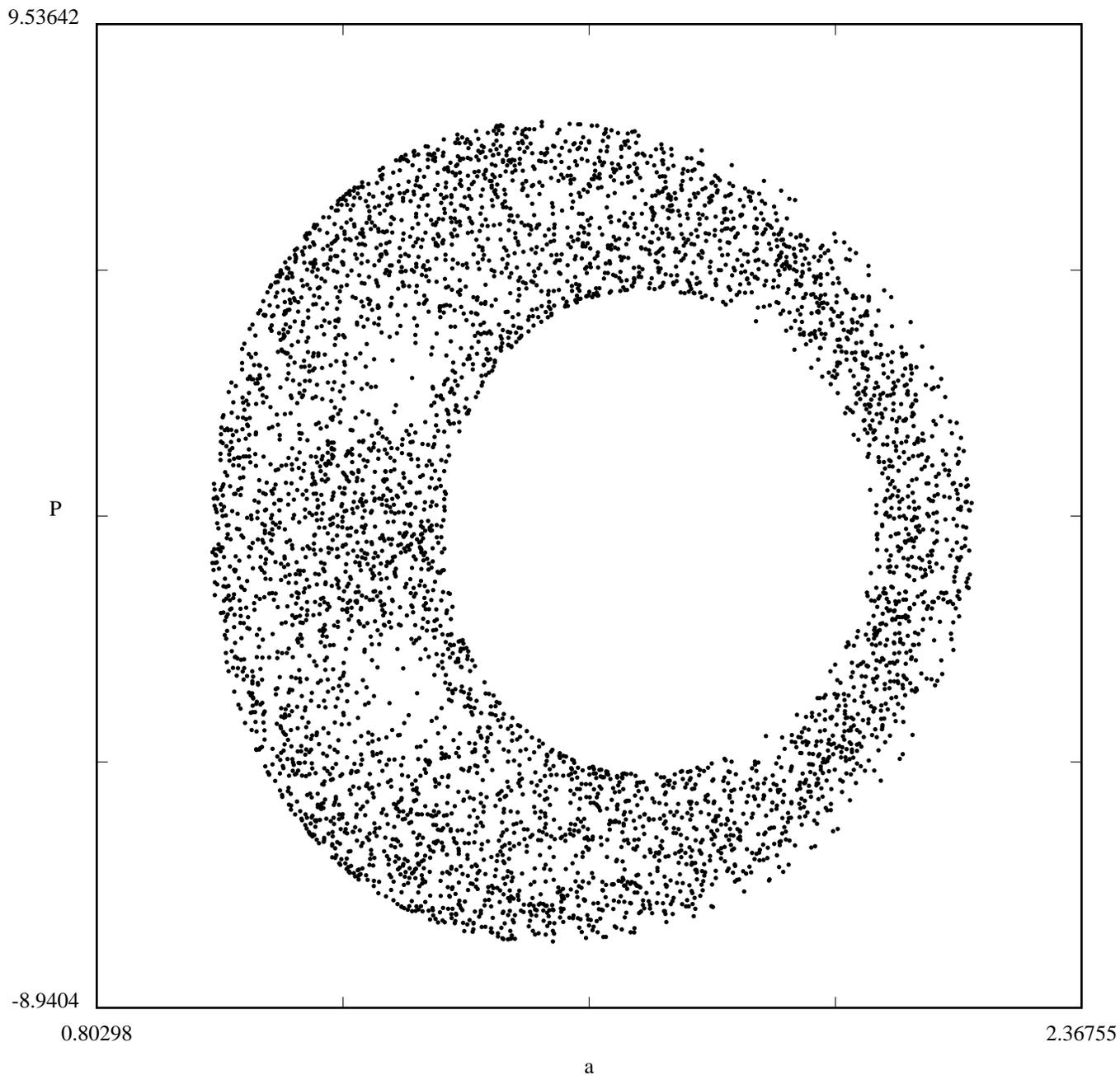
\end{document}